\documentclass[useAMS,usenatbib]{mn2e}
\usepackage{graphicx}
\usepackage{epstopdf}
\usepackage{natbib}
\usepackage{aas_macros}
\usepackage{color}
\usepackage{subfig}
\title{{\bf Some statistical remarks on the Giant GRB Ring}}

\author[L. G. Bal\'azs et al.]{Lajos G. Bal\'azs$^{1,2}$\thanks{E-mail,
balazs@konkoly.hu},L\'{\i}dia Rejt\H{o}$^{3,4}$, G\'abor
Tusn\'ady$^{4}$ \\
\\
$^{1}$MTA CSFK Konkoly Observatory, Konkoly-Thege M. \'ut 13-17, Budapest, 1121, Hungary\\
$^{2}$Department of Aastrojnomy, E\"otv\"os University,
P\'azm\'any P\'eter s\'et\'any 1/A,
Budapest,1117, Hungary\\
$^{3}$ Department of Applied Economics and Statistics, University
of Delaware, Newark, DE 19716, USA \\
$^{4}$ Alfr\'ed R\'enyi Mathematical Institute of the Hungarian
Academy of Sciences,  Budapest, P.O.Box 127, Hungary}

\begin{document}
\date{}
\pagerange{\pageref{firstpage}--\pageref{lastpage}} \pubyear{}
\maketitle \label{firstpage}

\begin{abstract}
We studied some statistical properties of the spatial point
process displayed by GRBs of known redshift. To find ring like
point patterns we developed an algorithm and defined  parameters
to characterize the level of compactness and regularity of the
rings found in this procedure. Applying this algorithm to the GRB
sample we identified three more ring like point patterns.
Although, they had the same regularity but much less level of
compactness than the original GRB ring. Assuming a stochastic
independence of the angular and radial positions of the GRBs we
obtained 1502 additional samples, altogether 542222 data points,
by bootstrapping the original one. None of these data points
participated in  rings having similar level of compactness and
regularity as the original one. Using an appropriate kernel we
estimated the joint probability density of the angular and radial
variables of the GRBs. Performing MCMC simulations we obtained
1502 new samples, altogether 542222 data points. Among these data
points only three represented ring like patterns having similar
parameters as the original one. By defining a new statistical
variable we tested the independence of the angular and radial
variables of the GRBs. We concluded that despite the existence of
local irregularities in the GRBs' spatial distribution (e.g. the
GGR) one can not reject the Cosmological Principle, based on their
spatial distribution as a whole. We pointed out the large scale
spatial pattern of the GRB activity does not necessarily reflects
the large scale distribution of the cosmic matter.

\end{abstract}


\section{Introduction}

In a recent paper \cite{Bal2015}  reported the discovery of a
Giant GRB Ring (GGR) consisting of 9 objects in the $0.78 < z <
0.86$ redshift range. The mean angular size of the feature is
$36^o$ corresponding to 1720 $Mpc$ in a comoving reference frame.
Voids surrounded by filaments are typical ingredients in the
cosmic matter distribution. Their characteristic size, however, is
at least an order of magnitude smaller than  that of the ring
\citep{Fri1995,Ein1997,Suh2011,Ara2013}.

GRBs are extremely energetic transients and distribute nearly
uniformly on the sky \citep[and references therein]{Bri1996}.
There are evidences, however, that the isotropy is valid only for
the long GRBs ($T_{90} > 10 \, s$). Nevertheless, it is not the
case at the short ($T_{90} < 2\, s$ ) and intermediate ($2 <
T_{90} < 10\, s$) duration bursts
\citep{Bal1998,Bal1999,Cli1999,Mes2000,Lit2001,Mag2003,Cli2005,Tan2005,Vav2008,Tar2015}.
Due to their immense intrinsic brightness they can be seen at very
large cosmological distances and sofar they are the only observed
objects sampling the matter distribution of the Universe as a
whole, in particular testing the validity of the cosmological
principle ($CP$) \citep{Ell1975}.

The original intention of Bal\'azs et al. was to test $CP$ and the
discovery of the GGR was only a byproduct. They pointed out that
the $f_{obs}(l,b,z)$ joint probability distribution of the $l,b$
angular coordinates and the $z$ redshift can be factorized into
$g(l,b)\times f_{intr}(z)$, if $CP$ is valid. Testing this
hypothesis they found, unexpectedly the GGR.

Assuming that $CP$ is valid there is an estimated transition scale
of about $370 \, Mpc$  and the size of all the existing structures
does not exceed it \citep{Yad2010}. Recently, large quasar groups
(LQG) were reported significantly exceeding this size
\citep{Clo2013}. The largest structure (3 Gpc in diameter in a
comoving reference frame) reported sofar is the enhancement of
GRBs spatial density was reported by \cite{Hor2014} and
\cite{Hor2015}.

Some concerns were raised , however, on  these features as real
physical objects. According to those concerns these features are
to big to be causally connected. \cite{Ein2016}  concluded that
the LQGs found in the quasar space distribution can be
reconstructed also from random samples making use a friend of
friend (FoF) algorithm. Based on the GRBs' detected by the Swift
satellite and have measured redshift \cite{Ukw2016} did not find
any clustering and deviation from the $CP$.

Making use the Metropolis-Hastings algorithm and the spatial
density distribution of the dark matter, as predicted by the MXXL
simulation \citep{Ang2012} Bal\'azs et al. made extensive studies
to find giant ringlike features, without any success. They found,
that the largest scale of the deviation from the $CP$ is $280 \,
Mpc$ corresponding to the result of \cite{Par2012} in the Horizon2
simulation \citep{Kim2011}.

Assuming a linear relationship between the  cosmic barionic matter
spatial and GRB number density Bal\'azs et al. calculated the mass
of the giant ring and  obtained a value of $1 \times 10^{18} \,
M_{\odot}$ which still represents an overdensity of a factor of 10
suggesting the ring mass is in the range of $10^{17}-10^{18} \,
M_{\odot}$ depending on the fraction of GRB progenitors in the
stellar mass distribution.

Bal\'azs et al. discussed also the possibility that the ring is a
projection of a spherical shell. In this case to get the observed
properties of the ring one have to assume that there was a period
of $2 \times 10^8 \, years$ of enchained GRB activity in the hosts
along the shell.

The estimated mass of the ring significantly depends on the
assumption of the GRB activity in their hosts displaying it. There
are two extrems of this activity: linear relationship on the
barionic spatial density or the GRB frequency is higher along the
ring but the matter density is the same as in the field. In the
latter case the ring like feature has nothing to do with the
spatial matter density and its existence does not violate the
$CP$.

Regular features in GRBs' spatial distribution, consequently, do
not necessarily violate the homogeneity and isotropy of the total
large scale matter distribution, i.e. the $CP$. Large scale
spatial patterns in the star forming  and in this case in GRB
activity which do not follow necessarily the general spatial
matter distribution can not be ruled out.

Since Bal\'azs et al. were looking for enhancements in the spatial
number density of GRBs and not for regular patterns  we try in the
following to find further ring like features in the same data set.
We repeat this procedure also on completely random samples in
order to find their significance.

After attempting to get further regular features in the data we
test the independence of the angular and redshift distribution. We
followed the way suggested by Bal\'azs et al. and a direct one to
obtain significance with an alternative method.

\section{Search for ring like features}

We intended to make a comparison with the results of Bal\'azs et
al. therefore we used in the following the same data set
consisting of 361 GRBs.

\subsection{Mathematical background}
\label{maba}
 \noindent A possible general procedure is the
following:

\begin{itemize}
\item[a)] Find an appropriate statistic which minima results in
the very nine points found by Bal\'azs et al,
 \item[b)] determine the distribution of the statistic and get
 significance of the features found in this procedure.
\end{itemize}

Scrutinizing the feature of the nine points a possible statistic
is the squared norm error of a circle in 3D which plane is
orthogonal to the line between the centre of the circle and the
origin (the observer). Actually the covering sphere of the nine
points does not contain any other points. This property might be
ensured by choosing first an aspirant point in space and next
choosing the nine nearest points to the centre among data points.

The  GGR is formed by nine points which are flanked by other eight
points according to the chronological order, thus the smallest
block containing  the ring consist of $m=17$ points. Let $n$ be
the sample size($n=361$ in our case) and $t$ be a fixed number
between 1 and $n-m+1$, i.e.   $1\leq t \leq n-m+1$ fixed.  Set

\begin{equation}
r_t^2 = \frac{1}{m} \sum_{i=0}^{m-1}  \parallel x_{t+i}
\parallel^2
   \qquad\qquad\qquad
\end{equation}

\noindent where $x_j$-s are  the vectors pointing to the $j$th
sample element,  $(1\leq j \leq n)$ from the observer. Let us
consider a sphere with radius $r_t$.
 Let $y$ be an arbitrary point  on  the sphere with radius $r_t.$  Let us denote by
 $s_i$  $(i=1,\ldots, m)$ the distances between $y$ and the above
 $x_t, x_{t+1},\ldots, x_{t+m-1}$.  Thus

 \begin{equation}
 s_i = \parallel x_{t+i-1} - y \parallel    \qquad i=1,\ldots, m. \qquad\qquad\qquad
 \end{equation}

 \noindent
 Let us order increasingly the  above $s_i$ distances.  The ordered sample is denoted by
 $z_1 \leq z_2 \leq \ldots \leq z_m$.

 Notice that the $s$-s and of course the $z$-s  for the fixed $t$ depend on
 the chosen point $y$ of the sphere with radius $r_t$ but the dependence is not
represented in the notation. Let us consider
 $z_1\leq z_2\leq \ldots \leq z_k$   and  set    $b(y)= z_k-z_1, c(y)=z_k.$
First we seek the minima of $b(y)$ in $y$ for fixed $t$:

\begin{equation}\label{rsize}
R_t = \inf_{\parallel y\parallel=r_t} b(y), \qquad\qquad\qquad
\end{equation}

\noindent and set $C_t$ for the flanking $c(y)$ i.e. $C_t$ denotes
the value of $c(y)$ corresponding to the very $y$ minimizing
$b(y)$. The statistic $R_t$ measures the resemblance to a ring  of
the best  $k=9$ points among the investigated $m=17$ points. In
the second step of the algorithm we take the minimum in $t$ of the
statistic $R_t$:

 \begin{equation}
 R= \min_{1\leq t \leq n-m+1} R_t  \qquad\qquad\qquad ,
 \end{equation}

\noindent $R$ measures the property having rings in the entire
data set.

Finally we defined the level of concentration of the features
found by the algorithm for all $t$. The area of a circle with
radius $r_t$ is $2\pi r_t^2$, that with radius $C_t$ is $2\pi
C_t^2$. The ratio

 \begin{equation}\label{clev}
 \rho_t = \frac{2\pi r_t^2}{2\pi C_t^2}
 \end{equation}

\noindent measures the {\em level of concentration} of the
features obtained. Scrutinizing the whole sequence $\rho_t$ it
turned out that the GGR is not the one which minimises $R_t$ but
it is the one which maximizes $\rho_t$.

\subsection{Searching rings in real data}
\label{reda}

 We made a run of the algorithm described in Section
\ref{maba} at first on real data consisting of 361 GRBs. We sorted
the objects according  to the radial distance and formed groups of
subsequent 17 GRBs step by step at each elements of the sample.

By running the algorithm we assigned a pattern  consisting of 9
GRBs to each objects in the sample. Running the algorithm results
in an $R_t$ thickness of the annulus embedding the 9 points of the
pattern. The local minima in $R_t$ values may indicate ring like
patterns.

\begin{figure}
  \includegraphics[width=8cm]{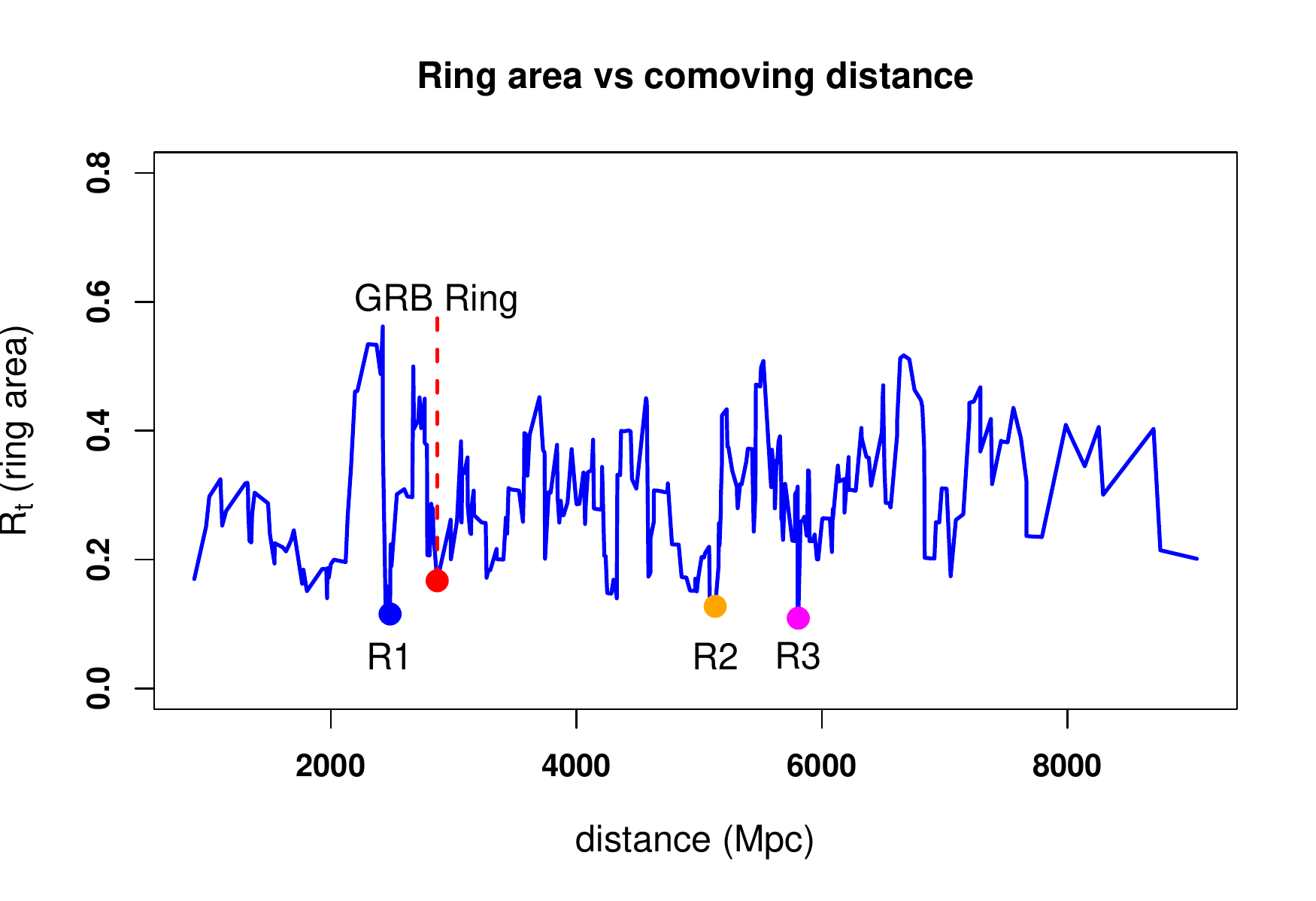}\\
  \caption{Dependence of the $R_t$ ring area, computed in Equation
  (\ref{rsize}),
  on the comoving distance. The minima deeper than that of the GGR
  may represent ring like features. The deepest three minima marked with
  colored dots and Labelled with $R1$, $R2$ and $R3$}\label{Radis}
\end{figure}

As one can infer from Figure \ref{Radis} the GGR is really lying
in a local minimum of $R_t$. There are, however, other minima
locating even deeper. We marked the three deepest ones with $R1,
\,R2$ and $R3$ in the Figure. The rings belonging to these minima
are displayed in Figure \ref{mufi}.

\begin{figure}
  \includegraphics[width=8cm]{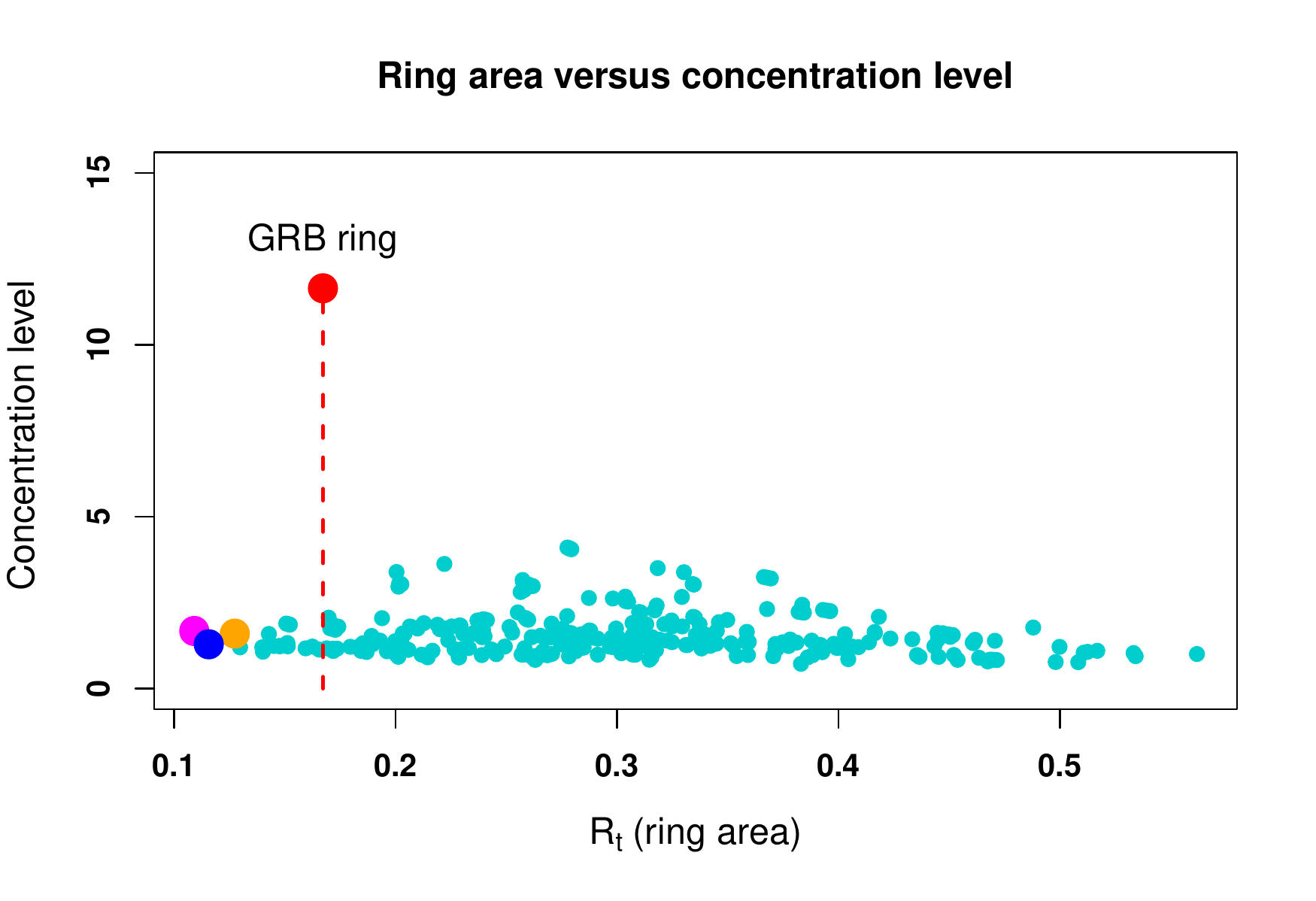}\\
  \caption{Dependence of the ring area concentration level (for definition
   see Equation (\ref{clev})) on the $R_t$ ring area (see Equation (\ref{rsize})). The
   colored dots (magenta, blue and orange) mark the minima labelled in Figure
   \ref{Radis}.}\label{Racon}
\end{figure}

At the end of Subsection \ref{maba} we defined a parameter the
level of concentration of a ring found in the procedure described
in this Subsection. In Figure \ref{Racon} we show the computed
level of concentration of each pattern found by the ring-searching
algorithm. One can see immediately that there is a marked
difference between the real ring and those found by simply
minimizing $R_t$.

Although, the $R1, \, R2$ and $R3$ objects show highly regular
circular pattern they  have much less level of concentration  than
the GGR. As Figure \ref{Racon} demonstrates it is quite an
isolated object in this respect.

\subsection{Searching rings in random samples}
\label{rada} We pointed out in Subsection \ref{reda} the GGR is
quite a unique object with respect to the level of concentration
within the studied sample of 361 GRBs. In the following we study
the problem of getting such a unique pattern quite accidentally.

\subsubsection{Searching in resampled data}
\label{resam}

Bal\'azs et al.  pointed out that in case of a valid  CP the joint
probability density of the angular and radial coordinates of the
GRBs can be factorized, i.e. it can be written as a product of the
radial and angular distribution. Supposing the validity of this
property a sample from the joint probability density of the
angular and radial distributions remains statistically invariant
if we reorder the radial distribution of GRBs, keeping the angular
positions unchanged.

\begin{figure}
  \includegraphics[width=8cm]{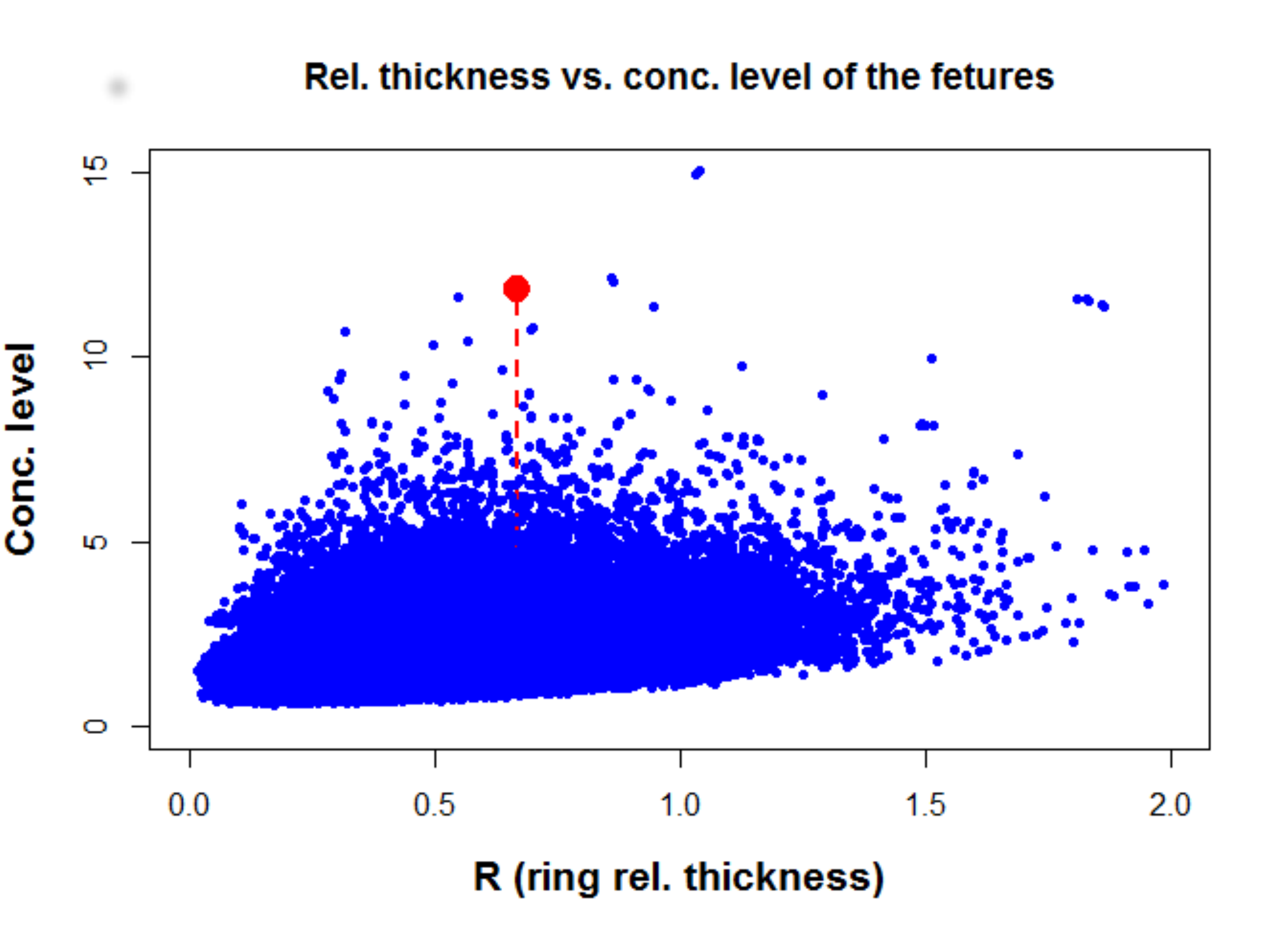}\\
  \caption{Result in searching rings in samples of resampled
  distances. Red dot marks the GGR.The figure consist of
  542222 dots {\bf obtained from 1502 simulations each consisting of
  361-17+1 objects ((See Subsection \ref{maba}). All of them  are
  representing a pattern of 9 points.} Note that there
  is no pattern having larger concentration and smaller relative
  thickness than those of the Giant GRB Ring. }\label{Resamp}
\end{figure}

To perform this reordering we invoked the \textcolor{red}{\tt
sample()} procedure of the R statistical
package\footnote{https://www.r-project.org/}. Making use 1502
times this procedure and combing the results with the unchanged
angular coordinates we obtained 1502 independent samples of 361
GRBs. We made run the algorithm described in Subsection \ref{maba}
in each of these samples, separately. The algorithm assigned a
pattern of 9 GRBs to each sample elements embedded in an annulus
of a width as given in Equation (\ref{rsize}). Having  $R_t$ we
get the relative width of the annulus dividing $R_t$ by the
internal radius of the annulus. We displayed this relative width
versus the level of concentration of the feature in  Figure
\ref{Resamp}.

As one can reveal from this Figure none of the simulated features
has higher level of concentration and smaller relative width of
the embedding annulus than the GGR.

\subsubsection{Searching in completely spatially random (CSR) data}
\label{mcmc} In  Subsection  \ref{resam} we reordered the
distances of the GRBs but left their angular position unchanged.
In this Subsection we simulate both the angular and radial
distribution as well, assuming again the statistical independence
of the angular and radial distribution, as we did it in
Subsububsection \ref{resam}.

Before performing the simulation to get the angular and radial
distributions we estimated the probability density functions of
the angular and radial data, separately. To get these probability
density functions  we did kernel smoothing of the observed sample
of GRBs. Having a sample of $(x_i, \, i=1,2,...,n)$ kernel
smoothing estimates the pdf at an arbitrary $x$ point by

\begin{equation}\label{kern}
    f(x)= c_0 \sum\limits^n_{i=1} K(x|x_i)
\end{equation}

\noindent where $c_0$ is a normalization constant and $K$ is am
appropriate  kernel.

 The angular positions of the GRBs
are distributed on a sphere. To do kernel smoothing one has to
define the functional form and characterizing length of this
procedure. For smoothing on spherically distributed data
\cite{Hal1987} suggest the functional form of $e^t$ where $t$ had
the form of $(ww_i-1)/h$. In this expression $ww_i$ is a scalar
product of unit vectors pointing to an arbitrary and the $i^{th}$
sample points on the sphere, respectively and $h$ is a smoothing
length. Easy to see that  $w=w_i$ gives  $t=0$ and $e^t=1$.

\begin{figure}
  \centering
  \includegraphics[width=7cm]{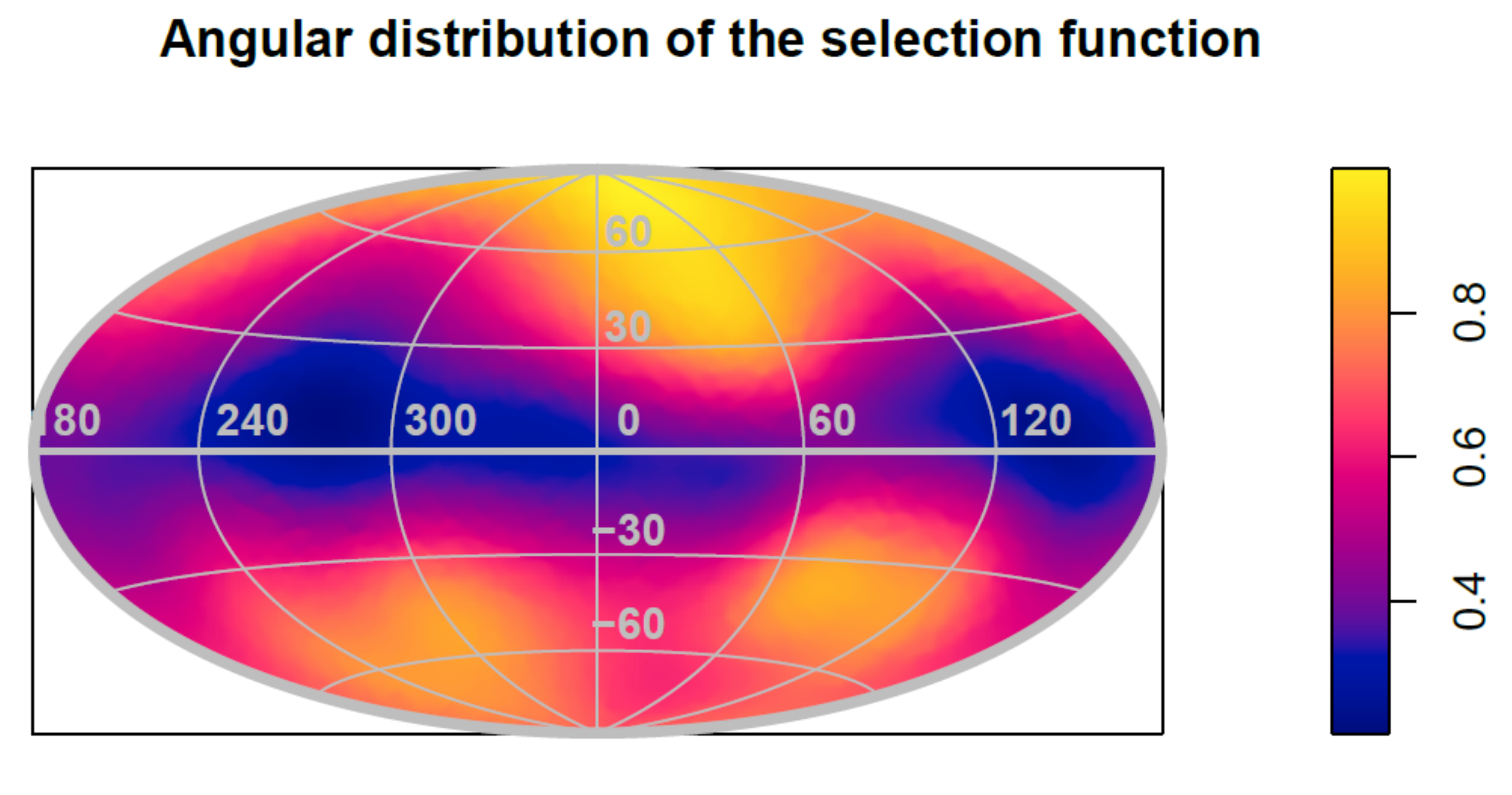}\\
 \caption{Color coded representation of the selection
function of Equation (\ref{kern}) in Aitoff projection of Galactic
coordinates. The dark strip along the equator is due to the
Galactic foreground extinction.}\label{self}
\end{figure}

For smoothing length we selected the largest value giving a pdf
which was statistically still compatible with the original sample.
For smoothing the radial data we used a kernel of a Gaussian form
\citep[see e.g.][]{Bag2016}. {\bf We showed the color coded
representation of the selection function in Figure \ref{self}}

Having in hand the pdf of both the angular and radial
distributions we simulated random samples  by making use   Markov
Chain Monte Carlo simulation (MCMC )realized by  the
Metropolis-Hastings algorithm \citep{Met1953,Has1970} available in
the \textcolor{red}{\tt metrop()} procedure in the $mcmc$ library
of the R statistical package. We made run the MCMC simulations for
getting angular and radial data, separately.

\begin{figure}
  \subfloat[]{\includegraphics[width=8cm]
  {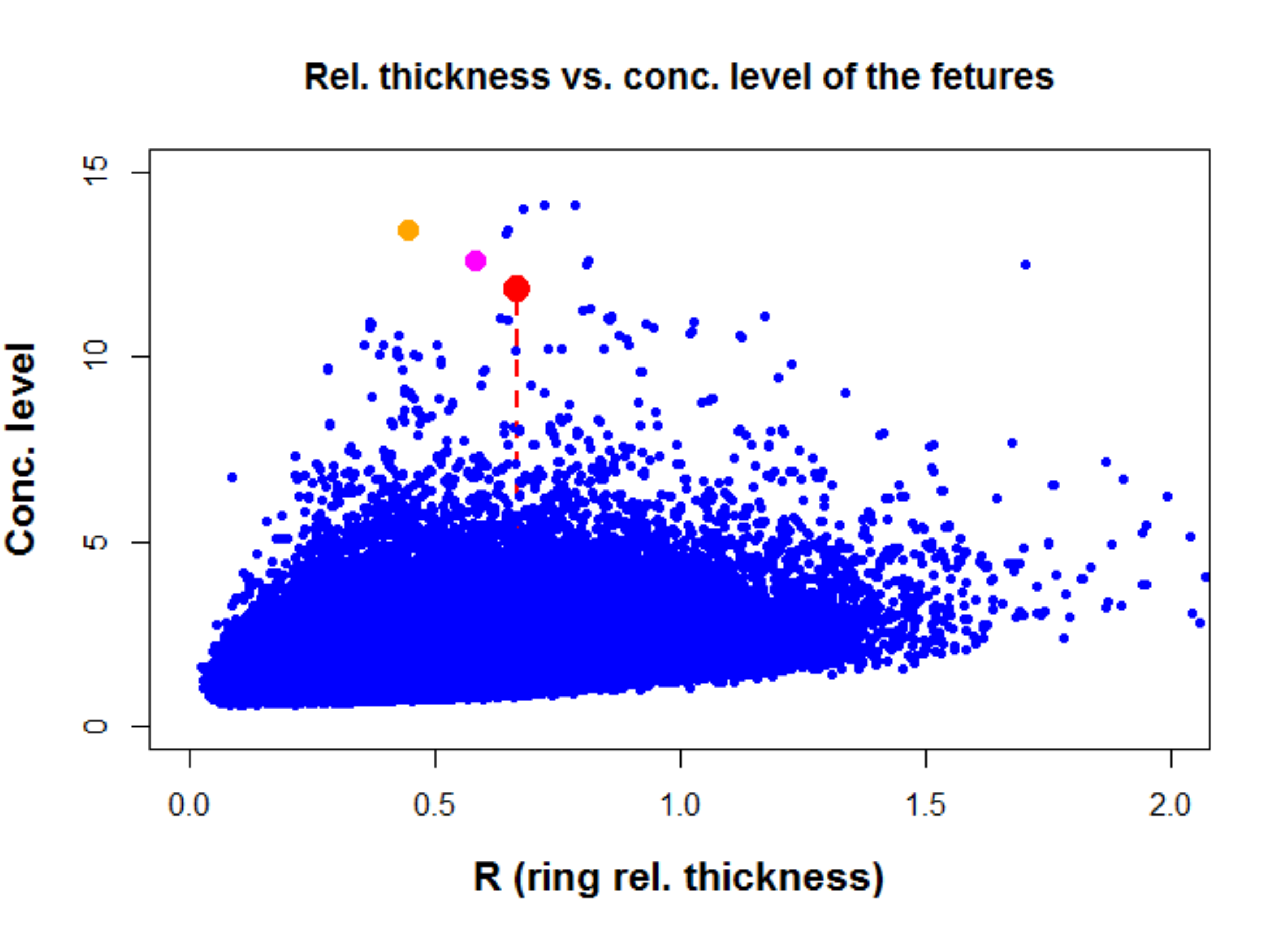}}\\
  \caption{ Scatter plot of relative thickness versus concentration level
  of patterns recognized  in random data. Red dot marks the GGR.
  Orange and magenta dots indicate simulated patterns
  having larger level of concentration and less thick embedding
  annulus than those of the Great GRB Ring. The number of simulations
  was 1502 each consisting of  {\bf 361-17+1 objects (see Subsection \ref{maba})
   resulting in 518190 dots}  in the Figure. Note that  there
  are only three points representing   patterns having  higher level
  of concentration and smaller relative  thickness than  that of  the GGR.}
\label{Racons}
\end{figure}

Performing this procedure 1502 times we obtained 1502 independent
samples each getting a size of 361, altogether  542222  objects in
total. By making run the algorithm of Subsection \ref{maba} we
assigned to all of these points a feature of 9 objects, each
having an embedding annulus of some level of concentration and
relative thickness according to the final part of \ref{maba} and
\ref{resam}. The result is displayed in Figure \ref{Racons}.

One may infer from Figure \ref{Racons} that only three points are
representing patterns of having higher level of concentration and
smaller relative with of the embedding annulus than that of the
GGR. This figure gives some hint for the probability of getting
such a ring like pattern fully accidentally (two of them are
marked with orange and magenta colors and displayed in Figure
\ref{simfi}).

\subsubsection{Searching in data of uniform angular distribution}
\label{unif}

 We showed in Subsubsections \ref{resam}  and  \ref{mcmc} that a
GRB ring having the same size and regularity than GGR is a low
probability phenomenon to get it only by chance. This conclusion
was based, however, on the resampled data and those obtained from
the MCMC simulation. In both cases the original angular
distribution of the objects were seriously biased by selection
effects as the exposure function of detecting GRBs, the
availability of the necessary telescope time for measuring
redshift, and the Galactic foreground extinction.

One may have concerns, therefore, that these selection effects may
seriously modify also the probabilities of getting ring like
features, such as the GGR. Namely, due to the angular
irregularities of the selection effects an existing ring like
feature can be recorded by parts only, and not detected by the
algorithm described in Subsection \ref{maba}.

In the Introduction we already mentioned that the long GRBs
distribute on the sky nearly uniformally. Most of the GRBs having
measured redshift belong to this group. Accepting that the true
angular distribution of GRBs is uniform the $f(x)$ surface density
defined in Equation (\ref{kern}) represent the actual selection
function inserted by the observations on the true angular
distribution.  Accepting this assumption one may estimate the size
of the unbiased sample from which the observed size was obtained.

Proceeding in this way we estimated the size of the true sampõle
of uniform angular distribution resulted in the observed one by
applying the $f(x)$ selection function. We obtained a size of
about 700 for the unbiased sample, in this way. We simulated a
sample with this size of uniform angular distribution. The
distribution of the GRB distances was obtained in the similar way
as in Subsubsection \ref{mcmc}.

\begin{figure}
  \includegraphics[width=8cm]{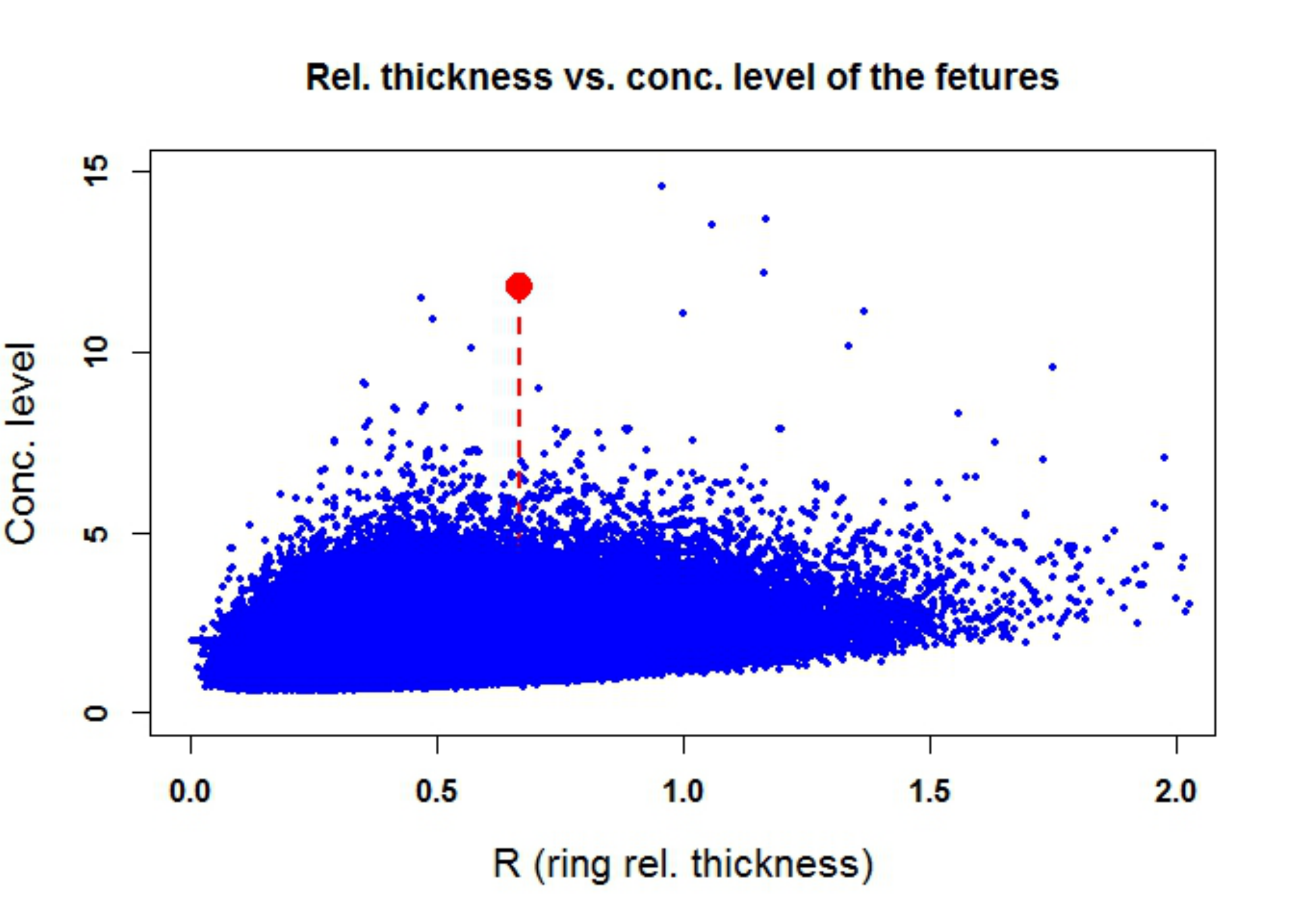}\\
  \caption{Result in searching rings in samples of uniform angular
  distribution. Red dot marks the GGR.The figure consist of
   1,029.555 dots obtained from 1503 simulations each consisting of 700-17+1
   objects (see Subsection \ref{maba}).  Each represents
   a pattern of 9 points. Note that there is no pattern having larger
   concentration and smaller relative thickness than those of the
   GGR.}\label{700ab}
\end{figure}

We made run the algorithm of Subsection \ref{maba} on the sample
obtained in this way and repeated this procedure 1503 times,
resulting 1,052.100 objects in total. The results is displayed in
Figure \ref{700ab}.  It clearly demonstrates that no simulated
features exists with higher level of concentration and regularity
than GGR.

In conclusion, getting a ring like feature, similar to GGR, has a
very low probability to find it purely accidentally, even in a
sample of unbiased uniform distribution.

\section{Testing independence}
\label{tein}

The original intention of Bal\'azs et al.  was to test the
validity of CP making use a sample of GRBs with known redshifts.
They pointed out that in case of a valid CP the joint pdf of the
observed angular and radial coordinates can be factorized into an
angular and radial part. The discovery of the GGR was only a
byproduct.

If the joint pdf can be factorized into an angular and radial part
it means  the angular and radial distributions of GRBs are
stochastically independent.

In simulating the joint angular and spatial distribution of GRBs
we assumed the stochastic independence between these distributions
and, consequently, the validity of factoring the joint pdf into a
radial and angular part. In the following we test the independence
of angular and radial distribution of GRBs, directly.

If $x_i$ is a vector pointing to the $i^{th}$ object in the sample
in the 3D comoving frame then $w_i$

\begin{equation}
w_i =\frac {x_i}{\parallel x_i \parallel}, \quad i=1,2.\ldots,n,
\end{equation}

\noindent points to its angular position. The $\parallel x_i
\parallel$ norm (length) is the radial distance of the object from the observer
in a comoving frame. Denoting with $\kappa_i$ the index of nearest
angular neighbor of $w_i$ on the sky, among GRBs in the sample we
get:

\begin{equation}
\parallel w_i - w_{\kappa_i }\parallel = \min_{t\neq i} \parallel w_i - w_t \parallel.
\end{equation}

\noindent Our statistic for testing independence of norms
(distances) and directions (angular positions) is

\begin{equation}
V = \sum_{i=1}^n (\varrho(\kappa_i) - \varrho (i))^2, \label{Vvar}
\end{equation}

\noindent where $\varrho(i)$ is the rank number of $\parallel x_i
\parallel $. The rank number is the serial number of an
 object after reordering the sample according to $\parallel x_i
\parallel $. If the angular and radial distributions are
stochastically independent the closeness in the angular position
does not imply a closeness in the radial distance of the objects.

Assuming the validity of the null hypothesis, i.e. the angular and
radial distributions are independent we can get the distribution
of the $V$ variable in Equation (\ref{Vvar}) using the samples
obtained from the simulations outlined in Subsubsections
\ref{resam} and \ref{mcmc}.

\begin{figure}
  \includegraphics[width=8cm]{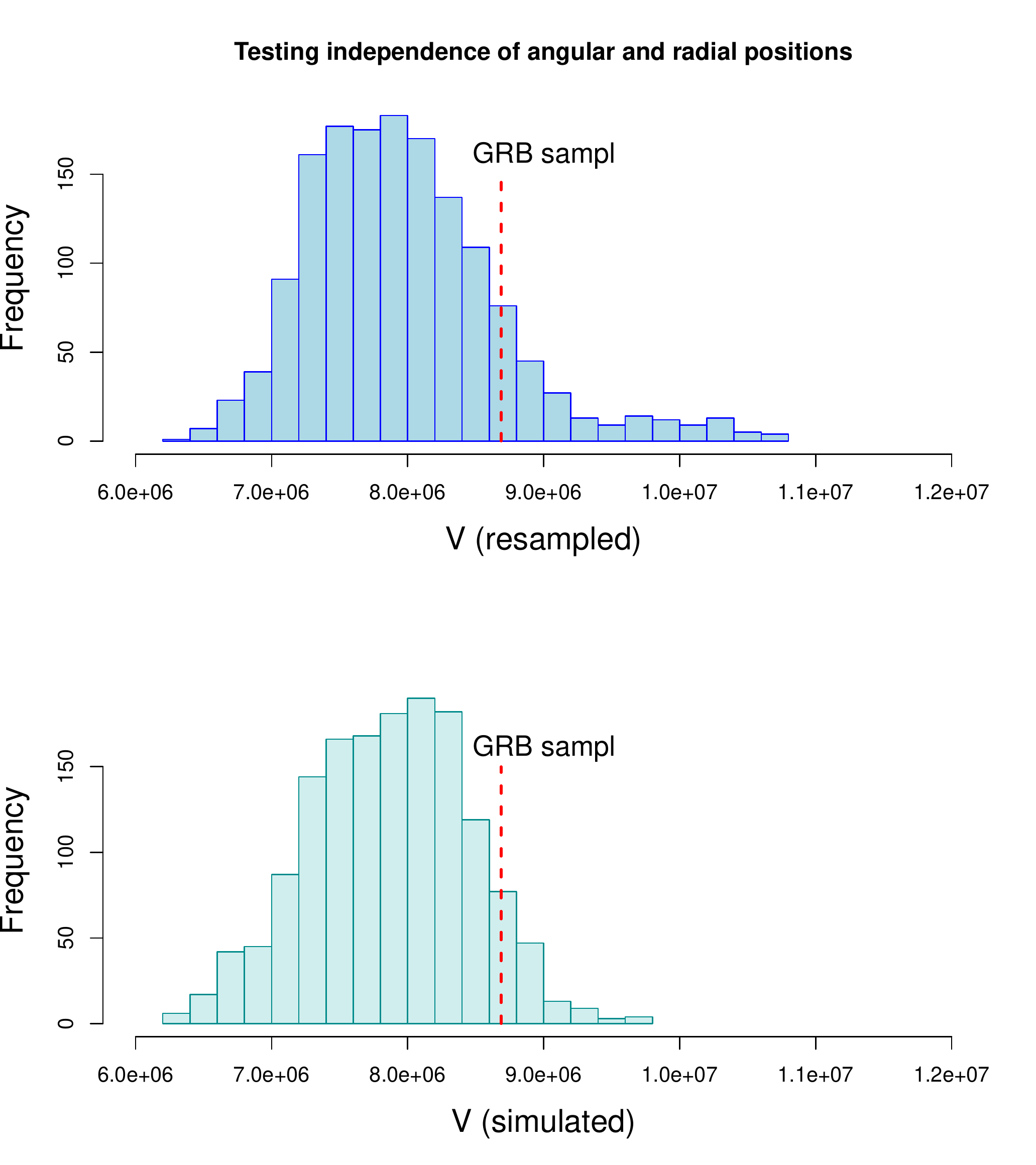}\\
  \caption{Distribution of the $V$variable defined in Equation
  (\ref{Vvar}). The upper panel shows the distribution obtained from
  the resampled data and the lower one from that of the
  completely random case. The red vertical dashed line indicates
  the $V$ value of the real GRB sample. The position of the real
  sample in the figures reveal that it does not contradicts
  to the null hypothesis, i.e. the angular and radial distribution
  of GRBs are statistically independent.}
  \label{angrad}
\end{figure}

The distributions of $V$ for the resampled and the completely
random  sample is given in Figure \ref{angrad} where the red
vertical dashed line indicates the value of the real GRB sample.
One can infer from comparing the histograms in the upper and lower
panel of the Figure that the distributions of $V$ in the resampled
and the completely random sample are not fully identical.

Already at the first glance, in Figure \ref{angrad} the maximum of
the upper histogram is a little bit shifted with respect to the
lower one. Furthermore, the upper histogram has a wing towards
higher $V$ values which is absent in the lower panel. Comparing
the two distributions by means of the Kolmogorov-Smirnov (KS) test
gives a probability of $p=5.48\times10^{-11}$ for the validity of
the null hypothesis, i.e. the upper and lower histograms of Figure
\ref{angrad} come from the same parent distribution.

There is a trivial explanation for the highly significant
difference between the $V$ distributions obtained from the
resampled and the completely random data. We may assume they have
different pdf,  although both can be factorized into an angular
and a radial part. Nevertheless, the random pdf was obtained by a
kernel procedure based on the true GRB data and keeping control on
the compatibility between the simulated and  the original
distribution. Therefore this trivial explanation can be excluded.

The different approach  in obtaining the samples can be the reason
for the stochastic difference between the resampled and the random
case. In getting the resampled data both the angular and radial
positions are identical with those of the original GRB sample. The
procedure changed only the order of the data. In the random case,
however, any position is eligible assuming we are still compatible
with the parent distribution obtained by the kernel smoothing. If
the original sample has some internal order which is invariant
against reordering it is still present after resampling unlike to
the random case.

It is easy to see that a clustering in the spatial distribution of
the objects results in systematically smaller values in $\varrho
(\kappa_i)-\varrho (i)$  differences  in Equation (\ref{Vvar}) and
the opposite is true in the case of voids. The presence of voids
is quite characteristic in the large scale distribution of cosmic
matter
\citep{Ein1980,Zel1982,Ick1984,Ick1991,Ein1997,Got2005,Ein2011,Ein2014}.
The resampling of the GRB data does not necessary destroy the void
structure and this may cause the wing of the larger $V$ values in
the upper panel of Figure \ref{angrad}.

In both cases, however, the factorization of the joint angular and
radial distribution is valid so the independency is true at this
statistical level.

Therefore, our result indicates that according to the distribution
of the $V$ variable there is no significant deviation from
independence based neither on the resampled nor on the complete
random samples.

\section{Discussion}

Testing the validity of CP is a basic problem of the observational
cosmology. Large scale deviation from the homogeneous isotropic
distribution of the cosmic matter casts a serious doubt on the
applicability of the  FLRW model for the Universe as a whole.

Until now the GRBs are the only objects sampling the space
distribution in the Universal matter as a whole. It is a problem,
however, that the number of GRBs with known redshift, i.e. with
spatial position, is very low (a few hundred, but steadily
increasing). Furthermore, it is also a problem wether their
spatial distribution represent a bias free sample of the global
distribution of cosmic matter including dark energy and matter.

GRBs are not physically similar objects. Traditionally, they were
divided into two basic classes: the short ($T_{90} < 2 \, sec$ and
long ($T_{90}>2 \, sec$) ones. There are some indications for an
intermediate duration group between them
\citep{Muk1998,Hor1998,Hor2002,Hor2006,Hor2008,Huj2009,Hor2009,
Hor2010,Uga2011,Tsu2014,Zit2015,Hor2016}. The vast majority of
GRBs having measured redshift belongs to the long group.
Observational evidences
\citep{Mes2006,Cha2007,Yuk2008,Kis2009,Wan2009,Ish2011,Wei2014}
connect the long GRBs to the star forming activity in the
underlaying host galaxy and theoretical arguments relate them to
the collapse of high mass (at least $25-30\, M_\odot$) stars into
a rotating black hole \citep[and the references therein]{Woo2006}.

\subsection{GRBs and large scale pattern of star forming rate}
\label{lssfr}

Because of the tight relationship of the GRBs, having measured
redshift, to the high mass star formation the observed spatial
distribution of these objects reflects some large scale/temporal
pattern of star forming activity and not necessarily  the
distribution of the cosmic matter as a whole.

As we pointed out above  the redshift/time distribution of GRB
activity versus the time dependence of cosmic star formation rate
are closely related. The large scale spatial distribution of the
star formation rate, however, is an open issue.

The large scale inhomogeneity in  the GRBs' spatial distribution
may have a close relationship with large scale spatial variation
of the star formation rate but not with the cosmic matter
distribution. Based on the Millenium simulation \citep{Spr2005}
\cite{Bal2015} presented evidences that the large scale spatial
distribution of the normal galaxies and those having high star
formation rate, are different.

Estimating the total stellar mass associated with the GGR Bal\'azs
et al. considered two extremes:

\begin{itemize}
 \item[a)]The general spatial stellar mass density
is the same in the field and in the ring's region and only the
star formation and consequently the GRB formation rate is higher
here.

\item[b)] There is a strict proportionality between the stellar
mass density and the number density of the progenitors.
\end{itemize}

For both estimates, one needs to know the local stellar density.
Proceeding in these ways they got a range of the mass of the ring
of $10^{17}$-$10^{18} \, M_\odot$.

Alternatively Bal\'azs et al. discuss the possibility that the
ring is actually a projection of a shell. The observed properties
is obtained if the ring is only a temporary configuration with an
estimated lifetime of $2 \times 10^8 \, years$. The estimated mass
of the whole structure is approximately an order of magnitude
greater: i-.e. $10^{18}$ - $10^{19} \, M_\odot$.

The masses in the above estimations consists of stars only.  In
the canonical $\Lambda CDM$ model, however, it is only a small
fraction of the total one. Unless, the GGR  was resulted in an
enhancement of starforming activity in the host galaxies and the
mass density inside is the same as in the environment it may
represent also a significant excess of the general matter density

The large amount of excess gravitating matter would have an
imprint on the cosmic microwave background  (CMB)
\citep{Cru2008,Gen2008,Mas2009,Das2009,Pad2009,Mas2009,Sol2010,Gra2010,
Bre2010,Chi2012,Fer2013,Kov2015,Kov2016} due to the integrated
Sachs-Wolfe effect  \citep{Sac1967}. A shell would have a ring on
the CMB.  Since  no such effect was observed in our case the ring
may be a large scale pattern of increased star forming activity
and not necessarily a density enhancement.

\subsection{Cosmic rings and global topology of the Universe}

No matter how the GRBs' large scale spatial distribution relates
to the total mass density their coordinates represent a stochastic
point process in the comoving reference frame, from strictly
statistical point of view. We followed this approach in the
present work, independently of the  real physical processes ending
up in their appearance as cosmic transients.

Treating the space distribution of GRBs strictly as a spatial
stochastic point process, regardless of its genesis,  we
demonstrated in Subsections \ref{reda} and \ref{rada} that  a
ring-like feature, having the same level of concentration and
regularity, is a rare event getting it  purely by chance.
Therefore it is worth studying physical processes resulting in
ring like structures.

Cosmological N-body simulations
\citep{Fal1978,Aar1980,Efs1985,Ber1991,Bag1997,Bag2005,Spr2005,Kim2011,
Ang2012,Joy2013,Val2015,Gar2016}
 attempted to reproduce the large scale spatial distribution
of dark matter. The simulations resulted in a system of strings
and voids (displaying rings in 2D projections) having the maximal
size of about 150 Mpc, i.e an order of magnitude less size than
the GGR The largest structure obtained in the s Horizon 2
simulation \citep{Kim2011} had a size of about 250 Mpc
\citep{Par2012}. The number of the existing GRBs with known
redshift is more than an order of magnitude less to that would be
necessary to reveal this structure \citep{Bal2015}.

The characteristics of the cosmic web resulted in these N-body
simulations depend on the choice of the initial conditions. In
this context assuming the homogeneity of the initial conditions
seems to be consistent with the observable angular isotropy of the
CMB. Although, the large scale isotropy of the CMB is widely
accepted \citep{Pla2014,Pla2016} there are attempts to find large
scale deviations from it
\citep{Haj2003,Bas2006,Sou2006,Lew2008,Aic2010,Pei2010,Zha2012,Muk2014}.

Solutions of Einstein's equation specify only the local properties
of the cosmological space. It does not constrain, however,  its
global topology. Conventionally, one assumes that the space is
simply connected and the infinity can be a reality. Nevertheless,
in a multiply connected case the physical world can be finite and
the infinity is only a pretence \citep{Paa1971}.

As \cite{Cor1998} pointed out if the size of the physical world is
less than the space surrounded by the last scattering surface
(LSS), appearing us as the CMB, the real LSS is intersected in
circles by its clones in the multiply connected world and can be
observed. The $\alpha$ angular radius of these circles can be
obtained from

\begin{equation}\label{ring}
\alpha=\arccos(\frac{X}{2R_{LSS}}),
\end{equation}

\noindent where $X$ is the size of the real world, and $R_{LSS}$
is the radius of the LSS.

Obviously, $R_{LSS}$ can be Substituted by the radius of any
sphere dedicated by some physical phenomenon, i.e. by GRBs, in
Equation (\ref{ring}). Of course, the size of the real world has
to be less than this radius. The  celestial  position of the ring
obtained in this way is given by the global topology of the cosmic
space and not necessarily populated by any ÛGRB events. Even if
the GRBs would populate the real space completely randomly the
probability to find an evens along the ring is higher.

There were many attempts to identify ring like patterns in the CMB
\citep{Mot2008,Mot2010,Mos2011,Weh2011,Mot2011,Gom2016} the
results, however, were not conclusive. \cite{Kov2010} found a
unique direction in the CMB sky around which giant rings have an
anomalous mean temperature profile. The score of the ring is close
to the direction of the cosmic bulk flow
\citep{Kas2000,Kas2008,Kas2009,Kas2010,Kas2011,Kas2013}. They
estimated the significance of the giant rings at the $3\sigma$
level. The recent detailed analysis of the Planck data
\citep{Pla2016}, however, concluded: there is no sign for a
multiple connected topology in the CMB data.

\subsection{Cosmic rings and large scale cosmological perturbations}

Large scale deviations of the cosmic matter density from the
homogeneous and isotropic case are accompanied with
inhomogeneities in the gravitational space might have footprints
on the CMB. The opposite is not necessarily true. Inhomogeneity in
the CMB not necessarily means gravitational irregularities. A
reason for it could be the improper elimination of the Galactic
foreground. The effect of the Galactic foreground is frequency
dependent but that of the gravitational irregularities is
achromatic.

In Subsection \ref{lssfr} we have already mentioned  that a mass
anomaly of the size of the GGR may result in a spot in the CMB,
due to the integrated Sachs-Wolfe effect. Since there is no such a
significant signal the statistical properties of the CMB would
make an upper limit for the density of this concentration.

In this context one may put the question  a matter concentration
of such a size could be possible at all? Let us suppose a flat
Euclidean spacetime and a small amplitude  perturbation in the
linear regime. The line element of the perturbed spacetime can be
written \citep{Bar1980}  in the form of

\begin{equation}\label{lpert}
   ds^2 = a^2[(1+2 \Phi)d\eta^2+2B_\alpha x^\alpha d\eta -
   (1-2\Phi)\delta_{\alpha \beta}dx^\alpha x^\beta]
\end{equation}

\noindent where $a(\eta)$ is the scale factor; $\eta$ is the
conformal time; $x^\alpha, \, \alpha = 1, 2, 3$, stand for the
comoving coordinates. The function $\Phi(\eta,r)$ and the spatial
vector $B(\eta,r) \equiv (B1,B2,B3)$ describe the scalar and
vector perturbations, respectively. In the Newtonian weak field
approximation the $\Phi(\eta,r)$ scalar function plays the role of
the gravitational potential.

Solving this problem \cite{Eing2016}  found a $\lambda$
characteristic length of $\Phi$ with a current value of $\lambda_0
\approx 3700 \, Mpc$. It is worth noting that this range and
largest known cosmic structures \citep{Clo2013,Hor2014,Bal2015}
are of the same order and their sizes essentially exceed the
previously reported epoch-independent scale of homogeneity $ \sim
370\, Mpc$ \citep{Yad2010}.

We have already mentioned that a spatial enhancement in the GRB
activity does not necessarily mean the same in the underlying
general matter density. The GRB activity directly relates to the
formation rate of the high mass stars. Based on the Millenium
simulation \citep{Spr2005} \cite{Bal2015} demonstrated: the
spatial distribution of the  galaxies  does not follow that of
having high star formation rate, in general.

Collision between galaxies is an  important source of the enhanced
star formation activity \citep[and the references
therein]{San1996}. However, this does not mean that interacting
galaxies are necessarily starbursts. Triggering depends on many
factors, e.g. the specific merging geometry and the progenitor
galaxies' properties
\citep{Mih1996,Spr2000,Spri2005,Cox2006,Cox2008,DiM2007,DiM2008,Tor2012,Mor2015}.
The frequency of collisions depends on the square of the number
density of the objects participating in collisions.

Let us suppose a $\delta\nu$ first order perturbation in the
$\nu_0$ spatial number density, $\nu  = \nu_0 + \delta \nu$, the
frequency of collisions proportional to $\nu^2 \approx
\nu_0^2+2\delta\nu$. Consequently, the amplitude of the increase
of the collision frequency is higher with a factor of two  than
that of the $\delta\nu$ number density enhancement. Therefore it
may happen that one finds anomalies in the GRBs' spatial
distribution at some level of significance which is not shown in
other cosmic objects.

In conclusion, density perturbations having the characteristic
size of the GGR may exist. The identification its imprint in the
spatial distribution of the observed objects needs a statistically
significant signal due to their spatial density.  As to the GRBs'
spatial distribution the amplitude of this signal is at least a
factor of two   higher in case if the high star forming activity
is given by the galaxy  collisions. Consequently, anomalies in GRB
distribution can  exist which are not necessarily shown in other
cosmic objects. Further detailed observations are necessary to get
a satisfactory solution of this problem.

\subsection{GRBs' spatial distribution and validity of CP}

The CP is valid  if the distribution of the cosmic objects  in the
comoving reference frame is completely spatially random (CSR). In
the CSR case the probability of finding exactly $k$  objects
within the volume $V$  with event density $\nu$
 is

 \begin{equation}\label{CSR}
    P(k,\nu, V)= \frac{(V\nu)^kexp(-V\nu)}{k!} \, .
 \end{equation}

\noindent In the above equation the $\nu$ event density is
constant throughout the whole comoving space. Nevertheless, it is
not true in the case of the spatial distribution of the observed
GRBs.

Even if the spatial density of the barionic matter is constant in
the comoving reference frame the formation of cosmic objects is a
long lasting complex procedure and their spatial distribution
reflects the history of their formation. As a consequence the
spatial homogeneity was lost. It is also true for the spatial
distribution of GRBs. In contrast, the isotropy is still hold.

In the case of isotropy the $\nu$ spatial number density depends
only on the distance of the object to the observer but not on the
angular coordinates. Since the sum of Poissonian  distributions is
also Poissonian and the column density replaces $\nu$ and the area
the $V$ volume.

In the case of spatial isotropy the angular distribution of GRBs
would be uniform, if there were no observational selection
effects. It can be demonstrated that correcting to the effect of
exposure function the angular distribution of long GRBs ($T_{90}>2
\, s)$ is isotropic but it is not the case of the short  ($T_{90}<
2 \, s)$ and intermediate ($2 < T_{90}< 10 \, s)$ ones (see the
references listed in the Introduction). The reason for this result
is the much larger volume sampled by the observed long bursts than
the short ones.

Only a small fraction (a few hundred) of the known GRBs have
measured redshifts, i.e. known distances. Besides the exposure
function the GRB distribution of known redshift suffers from the
selection effect due to the availability of the necessary
telescope time and the extinction of the Galactic foreground.

In the case of isotropy the $w$ angular position of an object is
statistically independent of the $r$ distance. Consequently, the
joint probability density $f(w,r)$ of the angular  position and
radial distance can be factorized into an $g(w)$ angular and
$h(r)$ radial part, i.e. $f(w,r)=g(w)h(r)$ \citep{Bal2015}. They
also pointed out the selection effects due to the telescope time
and Galactic foreground extinction does not influence this
factorization. The factorization means a stochastic independence
of the angular and radial positions.

The stochastic independence is a necessary, but not sufficient,
requirement of the validity of the CP. We tested this independence
in Section \ref{tein} and concluded that the GRB actual sample of
known redshift did not contradict to assuming stochastic
independence between the angular and radial positions.

This conclusion seems to contradict to the existence of the GGR
having a characteristic size of 1720 Mpc significantly larger than
the 370 Mpc transition scale \citep{Yad2010}  to the valid CP. The
GGR, however, consists of only nine objects. According to
Subsubsection \ref{resam} resampling the original GRB sample 1502
times did not reveal  ring like features having at least the same
level of concentration and regularity. However,shown in Section
\ref{tein} the whole sample was still consistent with assuming a
stochastic independence between the angular and radial
coordinates.

We may conclude, the local anomalies (e.g. the existence of the
GGR) in the sample of the GRBs with known spatial position does
not allow to reject the CP with a sufficiently high level of
significance. Even if they did, GRBs represent only a tiny
fraction of the cosmic matter density. We can not reject CP with a
high level of certainty until it is supported by a notable
fraction of the total cosmic matter density. Quite recently a
study of the general distribution of the cosmic dark matter
revealed it distributes much more evenly than previously was
thought
 \citep{Hil2017}.

\section{Summary and conclusion}
\label{suco}

We studied some statistical properties of a stochastic point
process defined by the positions of GRBs with known redshift in
the comoving reference frame. The sample studied was identical
with that used by \cite{Bal2015}  for discovering the GGR.

We developed an algorithm, described in Subsection \ref{maba}, for
finding ring like point patterns in the sample. Since the GGR
consists of nine bursts, concretely we were looking rings of this
size. This choice excludes ring with less element but find all
consisting of at least nine elements. We defined a concentration
level and a measure of regularity to get a similar pattern as the
giant ring.

Applying this procedure we identified additional three rings
having better level of regularity, displayed in Figures~
\ref{Radis} and \ref{Racon}. One may infer from these Figures,
however, the additional three rings found in this procedure have
better level of regularity but they are far less concentrated than
GGR.

Assuming a stochastic independence between the angular and the
radial coordinates we generated 1502 bootstrapped samples from
that used by \cite{Bal2015}, making use the \textcolor{red}{\tt
sample()} procedure of the R statistical package. The bootstrapped
samples consisted of altogether 542222 data points but none of
them participated in a ring like pattern of similar concentration
and regularity than that of the original GGR.

Using an appropriate kernel we estimated the angular and radial
probability density functions. Assuming again the independence of
the angular and radial coordinates we generated random samples. We
made Markov Chain Monte Carlo simulations (MCMC )realized by  the
Metropolis-Hastings algorithm available in the \textcolor{red}{\tt
metrop()} procedure in the $mcmc$ library of the R statistical
package.

Similarly, as we did in the case of the bootstrapped sample we
simulated 1502 samples consisting of all together 542222 data
points. As one can infer from Figure \ref{Racons} out of these
points only three represented ring like point patterns having at
least the concentration level and regularity as of the GGR.

We studied the effect of the selection bias on the probability of
getting a GGR like feature purely accidentally. This selection
bias comes into being through the superposition of the exposure
function of detecting GRBs, the availability of the necessary
optical telescope time and the Galactic foreground extinction. We
simulated 1503 samples each having a size of 700, 1,052.100
objects in total. (This is the size which is resulted in the
observed sample after applying the selection bias). We concluded,
the combined effect of this selection bias does not change
significantly the probability of getting a GGR like feature only
by chance.

In the simulations given in Subsubsections \ref{resam}, \ref{mcmc}
and \ref{unif} we assumed the stochastic independency of the $w$
angular and $r$ radial coordinates. In Section \ref{tein} we
defined a $V$ test variable measuring the level of independence.
Based on the distributions obtained from the bootstrapped and MCMC
simulations we obtained an empirical distribution of the $V$
variable.

Since the long GRBs relate to the high mass star formation their
spatial distribution represents a large scale footprint of this
process. We pointed out in Subsection \ref{lssfr} the large scale
pattern of GRBs' spatial distribution does not necessarily follow
that  of the total mass in the Universe.

Ring like features in the distributions of some special cosmic
objects may indicate a multiply connected global topology of the
Universe. Based on the latest analysis of the Plank satellites
data, however, one can exclude such an explanation of the GGR.

A galaxy-galaxy collision is one of the  major sources of the
enhanced star forming activity and consequently of the GRB rate.
The number of collisions is proportional to the square of the
number density of galaxies. In the case of small amplitude
perturbations it gives a factor of two higher statistical signal
in the spatial number density of GRBs than the underlying matter
density in general. One needs further observations to uncover the
true relationship between the spatial number density of GRBs and
that of the underlying matter in general.

Comparing the simulated distributions with the $V$ value of the
real GRB sample(see Figure \ref{angrad}) we demonstrated that it
did not contradict to assuming stochastic independence between the
$w$ angular and $r$ radial coordinates also in the real case. This
result also implies that as a whole the GRB sample of known
redshifts  does not contradict to the CP.

\section*{Acknowledgements}
This work was supported by the OTKA grant NN 111016. {\bf We are
grateful to Dr. Jon Hakkila, the referee, for his advises and
recommendations}. The authors are indebted to N.M Arat\'o, Z.
Bagoly, I. Horv\'ath, I.I. R\'acz and L.V. T\'oth for valuable
comments and suggestions.

\bibliography{Giant_ring}

\appendix
\section[]{Examples of ring-like patterns }

\begin{figure*}
 \subfloat[Giant GRB Ring]{\includegraphics[width=8cm]{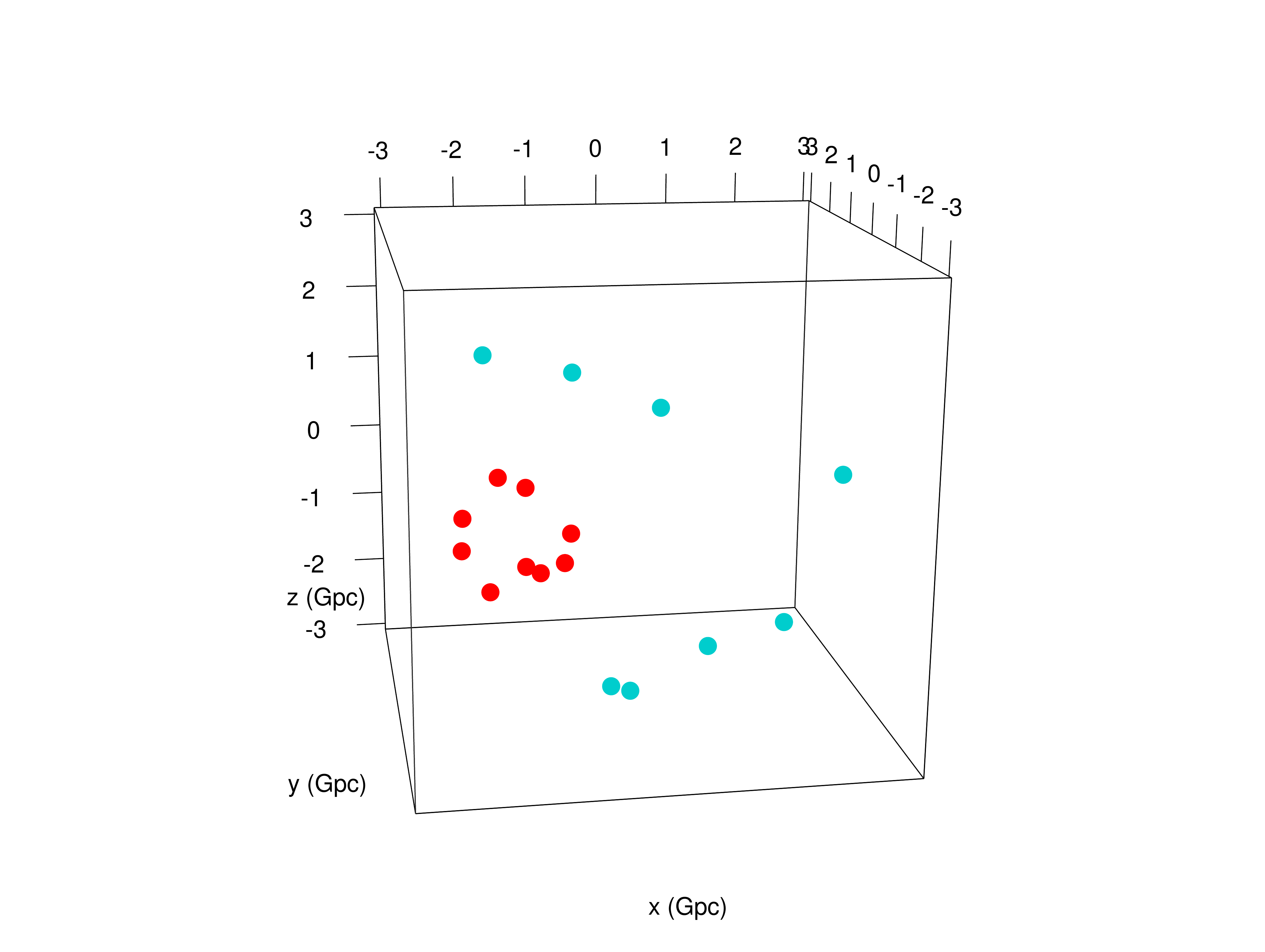}}
 \subfloat[R1 Ring]{\includegraphics[width=8cm]{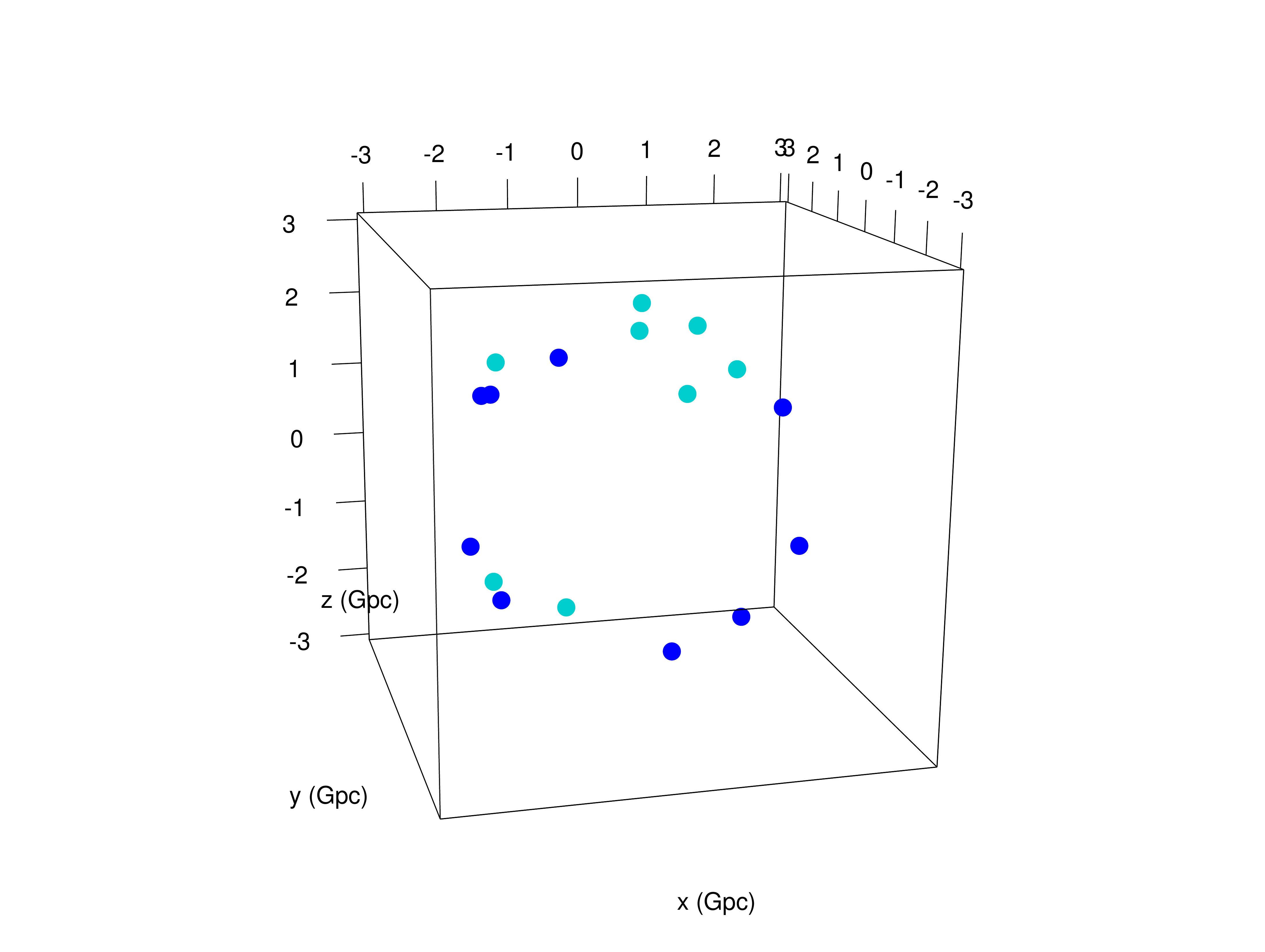}}\\
 \subfloat[R2
 Ring]{\includegraphics[width=6.0cm]{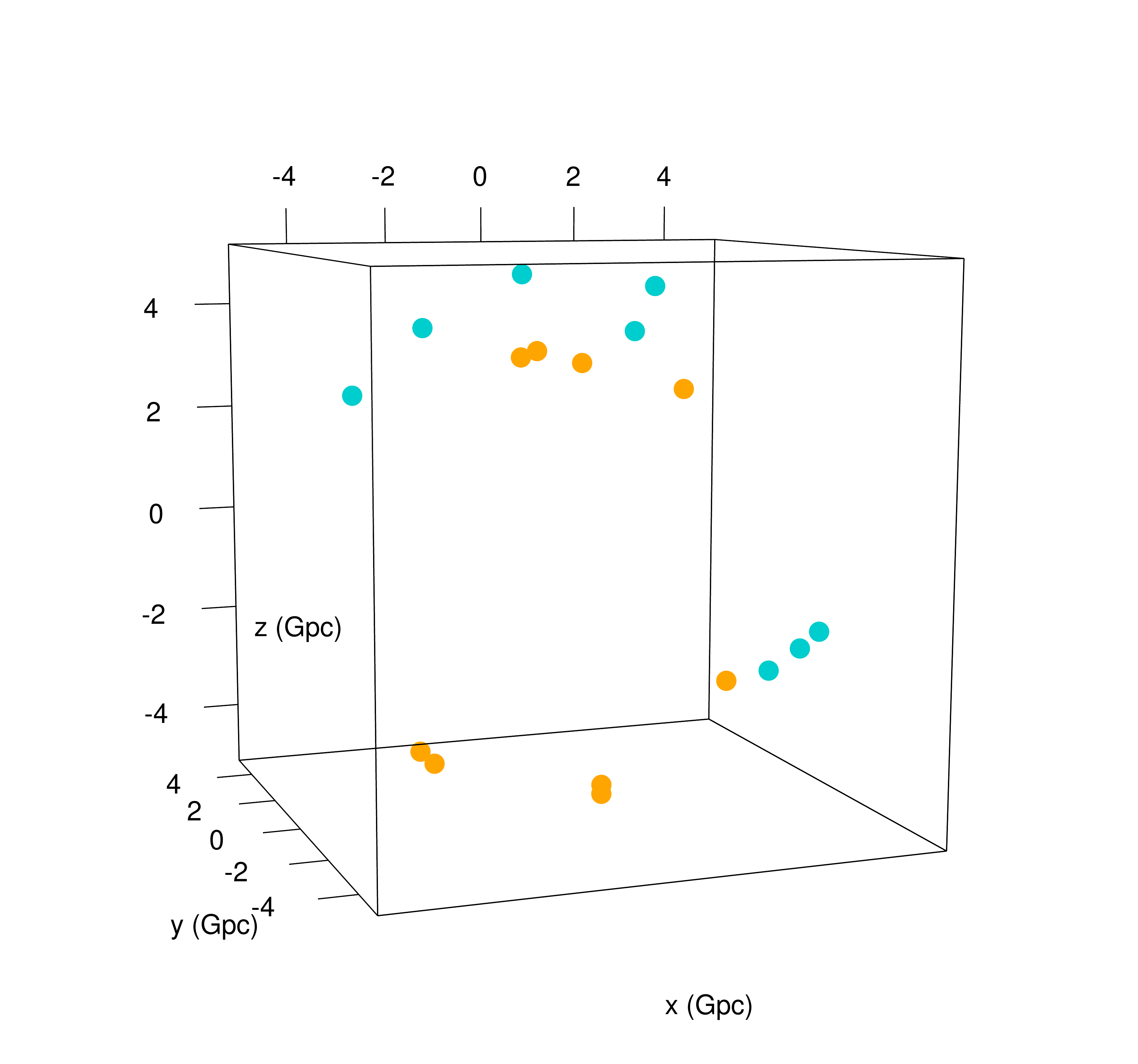}}\hspace{10mm}
 \subfloat[R3 Ring]{\includegraphics[width=6.0cm]{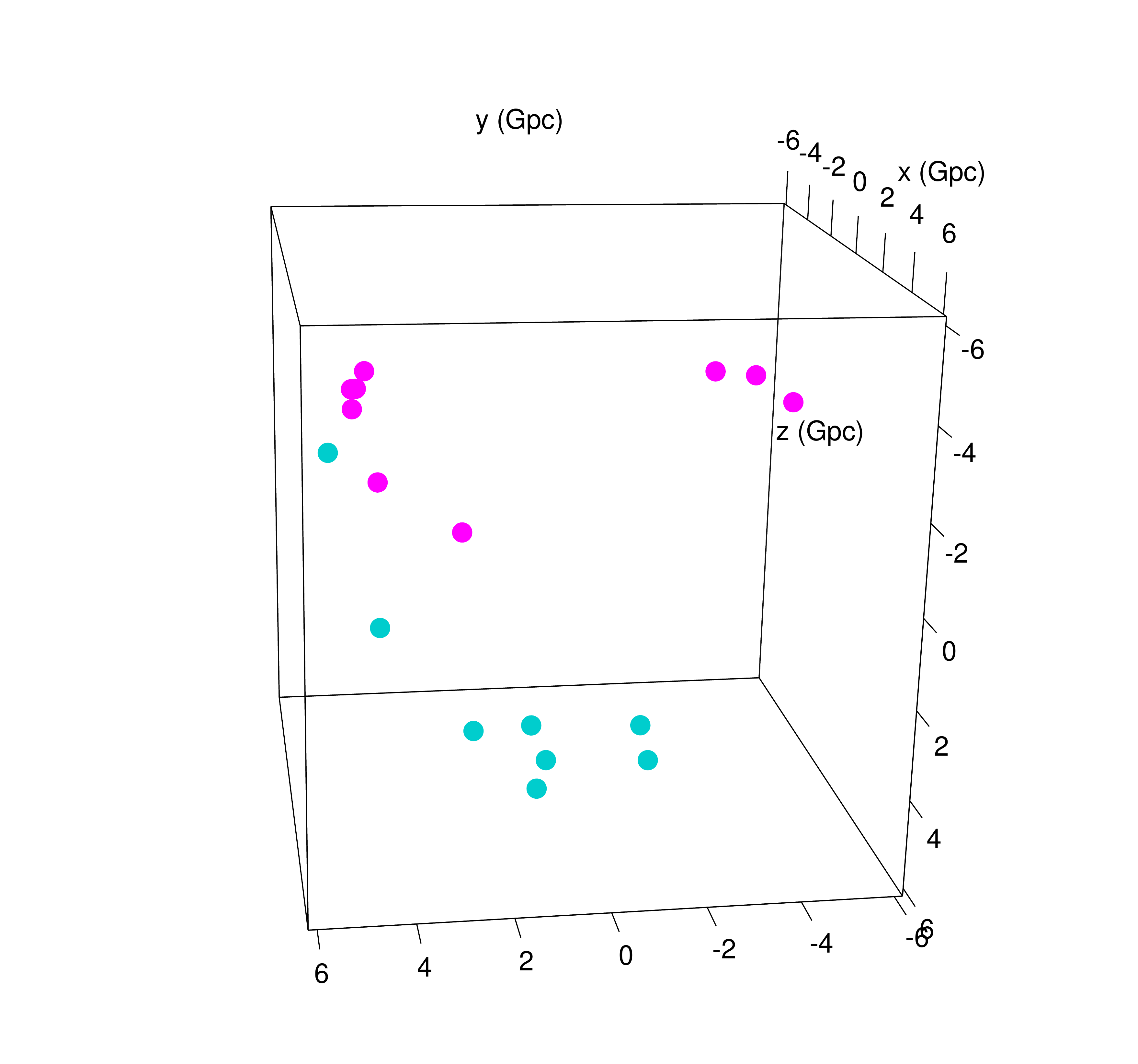}}
  \caption{3D plot of the Giant GRB ring found by Balázs et al.
  (2015) and those found by the procedure in Section \ref{maba} .
  The members of the rings are colored according to those
  in Figure~\ref{Radis}. The objects lying in the same distance range  but
  are nonmembers  marked with cyan color. The scale is given in
  $Gpc$.
  }\label{mufi}
\end{figure*}

\begin{figure*}
 \subfloat[Random Ring1]{\includegraphics[width=8cm]{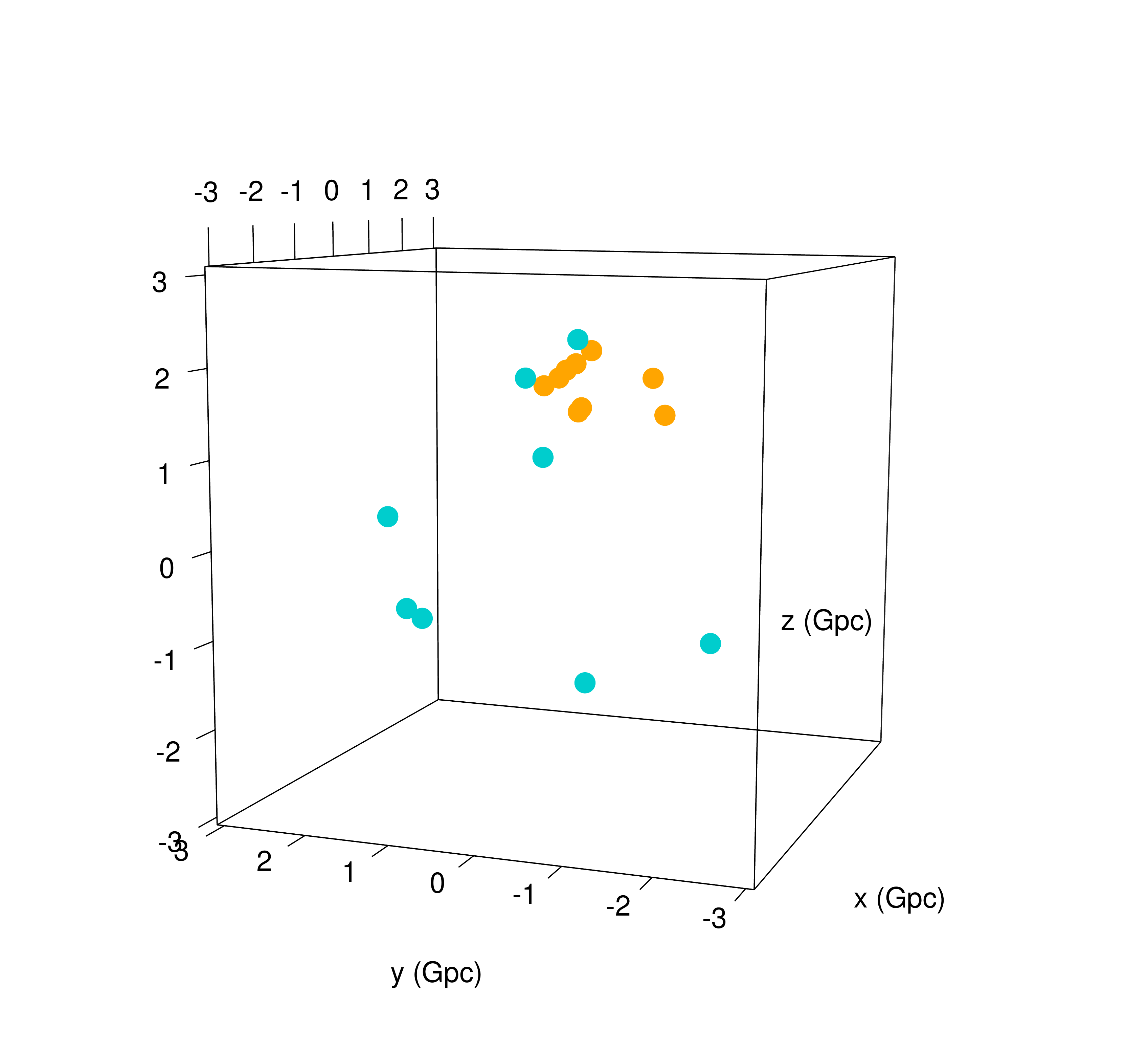}}
  \subfloat[Random Ring2]{\includegraphics[width=8cm]{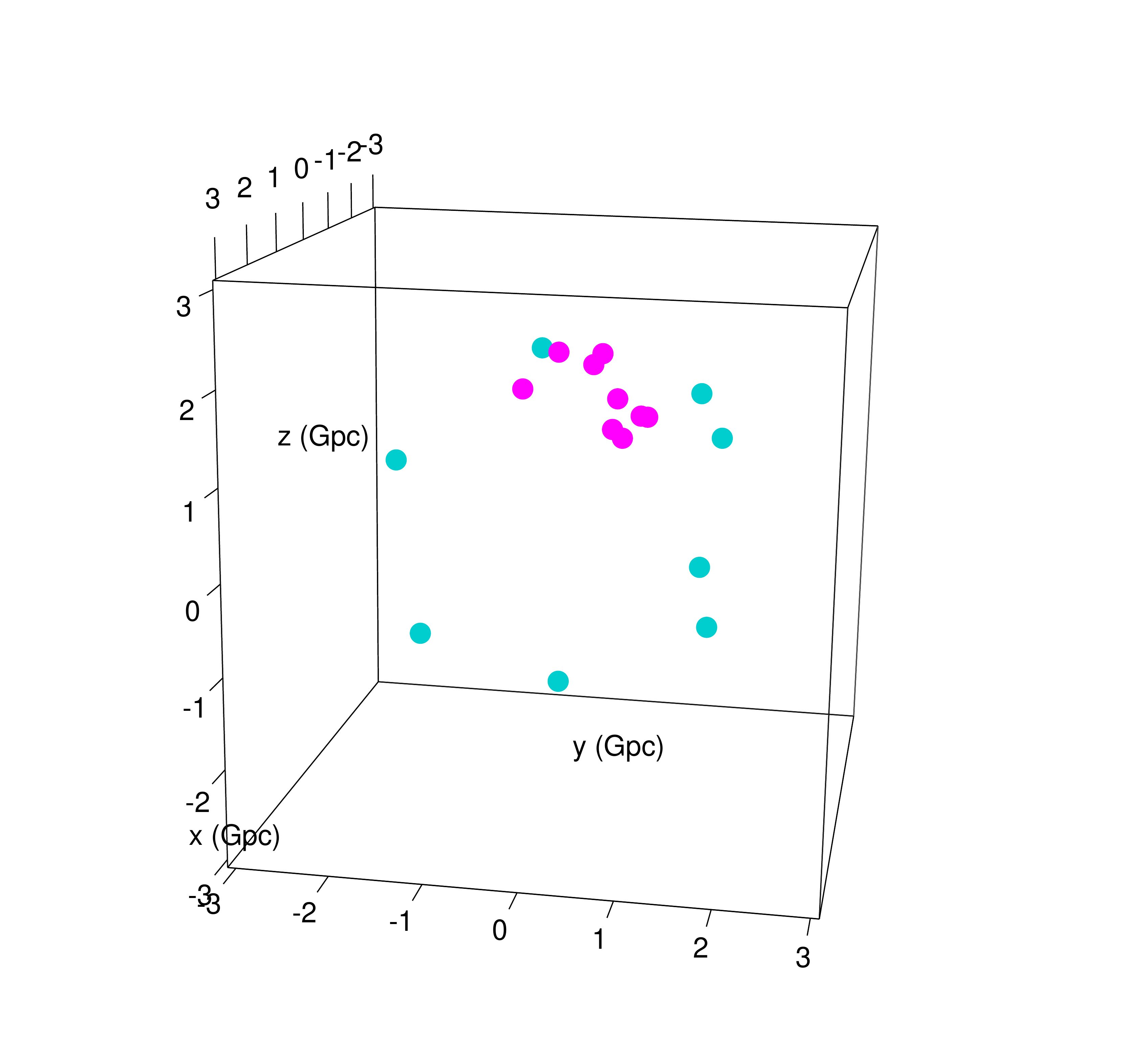}}
\caption{3D plot of rings obtained from random samples. (The
colors of the ring patterns correspond to those applied in Figure
\ref{Racons}).}
 \label{simfi}
 \end{figure*}

 \end{document}